\documentclass{elsart}%
\usepackage{amssymb}
\usepackage{amsfonts}
\usepackage{amsmath}
\usepackage{graphicx}%
\providecommand{\U}[1]{\protect\rule{.1in}{.1in}}
\begin{document}
%
\begin{frontmatter}%
%

\title{Insurance, Reinsurance and Dividend Payment}%
%

\author{Dan Goreac}%
%

\address
{Laboratoire de Mathématiques, Unité CNRS UMR 6205, Université de Bretagne Occidentale, 6, av. Victor LeGorgeu, B.P. 809, 29200 Brest cedex, France
\\Email: Dan.Goreac@univ-brest.fr\\Tel.: +33 (0)2 98 01 61 92\\Fax: +33 (0)2 98 01 67 90}%
%

\begin{abstract}
The aim of this paper is to introduce an insurance model allowing reinsurance
and dividend payment. Our model deals with several homogeneous contracts and
takes into account the legislation regarding the provisions to be justified by
the insurance companies. This translates into some restriction on the
(maximal) number of contracts the company is allowed to cover. We deal with a
controlled jump process in which one has free choice of retention level and
dividend amount. The value function is given as the maximized expected
discounted dividends. We prove that this value function is a viscosity
solution of some first-order Hamilton-Jacobi-Bellman variational inequality.
Moreover, a uniqueness result is provided.%
\end{abstract}%
%

%

\end{frontmatter}%

\section{Preliminaries}

A common problem of the insurance companies is to find a strategy allowing to
satisfy the claims appearing either from the insured parties as consequence to
specified peril or from the shareholders in terms of dividends. To reduce
their risks and protect themselves from very large losses, the companies
usually choose to pay some of the premiums to a third party. This process is
called reinsurance, and it commits the third party (the reinsurance company)
to cover a certain part of the claims. It is obvious that the insurance
company controls the contracts to be reinsured as well as the dividends to be
paid to the shareholders. These elements justify the framework of stochastic control.

This paper considers a utility function given as the maximized expected
discounted dividends. In the literature, this approach has been first used by
Jeanblanc, and Shiryaev (1995). In their model, the capital of an insurance
company is described with the help of a standard Brownian motion and the
dividend payment strategy is understood as control process. More precisely,
they deal with the following model
\[
dX_{t}=\mu dt+\sigma dW_{t}-dZ_{t},
\]
where $\mu$ and $\sigma$ are arbitrary constants, $W$ is a $1$-dimensional
standard Brownian motion and $Z$ is an adapted, non decreasing,
right-continuous process which represents the dividend payment strategy.

In Asmussen et al. (2000), a model concerning excess-of-loss reinsurance and
dividend payment has been studied. They use diffusion and proportional
reinsurance for their model. More exactly, they take as model for the capital
of the insurance company the process given by the following equation
\[
dX_{t}=a_{t}\left(  \mu dt+\sigma dW_{t}\right)  -dZ_{t},
\]
where $0\leq a_{t}\leq1$ stands for the retention level. In the case where the
rate of dividend pay-out is unrestricted, \ they characterize the value
function as the (classical) solution of some associated
Hamilton-Jacobi-Bellman equation.

The same problem is studied by Mnif, Sulem (2005), but the claims are
represented by a compound Poisson process. In their collective risk model, a
retention level is an adapted process $\alpha_{t}$ which specifies that, for a
claim $y$, the direct insurer covers $y\wedge\alpha_{t}$, while the
reinsurance company covers the remaining $\left(  y-\alpha_{t}\right)  ^{+}.$
They consider a single insurance contract and the reserve of the insurance
company satisfies%
\[
dX_{t}=p(\alpha_{t})dt-\int_{B}\left(  y\wedge\alpha_{t}\right)
\mu(dtdy)-dL_{t},
\]
where $\mu$ is the random measure associated to the compound Poisson process.
In the above equation, $p(\alpha_{t})$ is the actual premium of the insurance
company given the retention level $\alpha.$ The process $L$ describes the
pay-out of dividends for shareholders and it is an adapted, c\`{a}dl\`{a}g
process such that $L_{t}-L_{t-}\leq X_{t-}$ for all $t\geq0$. The value
function is defined as the maximized expected discounted dividends until the
ruin time $\tau,$%
\[
V(x)=\sup_{(u,L)}E\left[  \int_{0}^{\tau}e^{-rs}dL_{s}\right]  ;
\]
here $r$ is some positive discount factor. The authors proved that, under the
assumption that the value function satisfies the dynamic programming
principle, $V$ is a viscosity solution of the associated
Hamilton-Jacobi-Bellman variational inequality.

In the present paper we consider the problem of optimal reinsurance and
dividend pay-out with several insurance contracts. We will prove that in the
framework of the collective risk model, even if the invested initial capital
is arbitrarily small, one can expect a gain which exceeds an a priori fixed
positive constant. Indeed, this comes from the fact that, independently of its
initial capital, the model allows the insurance company to sell one contract.
However, as it is precised in section 2, in the case of insurance companies,
the codes of law impose that, at any time, these companies should be able to
justify enough resources to cover the obligations contracted towards their
clients. This condition imposes an upper limit for the number of contracts the
company can have. In the work we present here, several contracts are
considered. We obtain a stochastic differential equation with respect to a
random measure and introduce the utility for the shareholders as in Mnif,
Sulem (2005) to be the maximized discounted flow of dividends. We prove that
the value function is regular enough (enjoys the Lipschitz property) and
satisfies the associated Hamilton-Jacobi-Bellman Variational Inequality in the
viscosity sense. We also provide an uniqueness result for the viscosity
solution in the class of continuous functions of at most linear growth. We
emphasize that the limitation of the number of contracts which comes from the
codes of insurance, allows us to get the Lipschitz property of the value
function $V$. This property insures that an initial capital close to 0 will
induce a zero-expected gain (unlike the collective risk model). Moreover, in
this case, the dynamic programming principle follows in a standard way, while
it was only assumed by the authors of \cite{MS}.

The paper is organized as follows. In the first section we present a simple
example showing the limits of the collective risk model. The second section is
concerned with the insurance problem with several contracts. We introduce the
model, the basic assumptions and prove some elementary properties of the value
function $V.$ In the third section, we show that the value function is a
viscosity solution of the associated Hamilton-Jacobi-Bellman variational
inequality. The fourth section provides a comparison result which allows to
obtain the uniqueness of the viscosity solution for the given variational
inequality. A numerical example is given in the last section.

\section{The limits of the collective risk model. A counter example}

We consider the following special case of the collective risk model introduced
by Mnif, Sulem (2005). We assume that the claims are generated by a Poisson
process $N$ with intensity $1$ on a complete probability space $(\Omega
,\mathcal{F},P)$. We denote by $\left(  \mathcal{F}_{t}\right)  _{t\geq0}$ the
filtration generated by the random measure associated to $N,$ completed by the
family of $P$-null sets. Given an $\mathcal{F}_{t}-$adapted process
$\alpha_{t}\in\left[  0,1\right]  $ (retention level), the premium rate is%
\[
p(\alpha_{t})=k_{1}-k_{2}+\left(  1+k_{2}\right)  \alpha_{t},\text{ for all
}t\geq0,
\]
where $0\leq k_{1}\leq k_{2}$ are proportional factors. Moreover, if $L$
denotes the $F_{t}$-adapted process of cumulative dividends, then the reserve
of the insurance company satisfies the equation%
\[
X_{t}^{x,\alpha,L}=x+\int_{0}^{t}p(\alpha_{s})ds-N_{t}-\int_{0}^{t}dL_{u}.
\]
The process $L$ should be right-continuous, non-decreasing and such that
$L_{0-}=0$ and $L_{t}-L_{t-}\leq X_{t-}^{x,u,L}$ for all $t\geq0$. We
introduce the first jump time for the Poisson process $N$%
\[
\tau_{1}=\inf\left\{  t\geq0:N_{t}=1\right\}  .
\]
Obviously, $\tau_{1}$ is of exponential law with intensity $1$, and, in
particular,%
\[
P\left(  \tau_{1}>1\right)  =e^{-1}.
\]
If we consider the strategy $\left(  \alpha,L\right)  $ given by
\[
\left\{
\begin{array}
[c]{l}%
\alpha\equiv1,\\
L_{t}\left(  \omega\right)  =I_{\left\{  \tau_{1}>1\right\}  }(\omega
)I_{\left\{  t\geq1\right\}  }(t),
\end{array}
\right.
\]
then $\left(  \alpha,L\right)  $ is admissible and the ruin time%
\[
\tau^{x,\alpha,L}>1\text{ on }\left\{  \tau_{1}>1\right\}  .
\]
Indeed,
\[
X_{t}^{x,\alpha,L}=x+\left(  1+k_{1}\right)  t-N_{t}-\int_{0}^{t}dL_{u},
\]
and on $\left\{  \tau_{1}>1\right\}  $ we have that
\[
X_{t}^{x,\alpha,L}=x+\left(  1+k_{1}\right)  t,
\]
for all $t<1.$

It follows that
\[
V\left(  x\right)  \geq E\left[  \int_{0}^{\tau^{x,\alpha,L}}e^{-rt}%
dL_{t}\right]  \geq E\left[  e^{-r}I_{\left\{  \tau_{1}>1\right\}  }\right]
\geq e^{-\left(  r+1\right)  },
\]
for all $x>0.$ Obviously%
\[
V\left(  0+\right)  \geq e^{-\left(  r+1\right)  }>0.
\]
Therefore, investing an arbitrarily small capital in the insurance company, we
expect to gain more than $e^{-\left(  r+1\right)  }.$ This contradicts theorem
3.3 in \cite{MS}. This problem is due mainly to the fact that, independent of
the initial capital, the insurance company is allowed to hold one contract.

However, the insurance law requires that, at any moment, the companies should
be able to cover any liabilities that have been incurred on insurance
contracts as far as can be reasonably foreseen. Experience of similar claim
development trends is of particular relevance. Usually, the solvency margin is
computed with respect to both the premium rates and the average claim.
According to the current Solvency I prudence regime, "the life insurance
capital requirements are arrived at by multiplying a factor of 4\% to the
mathematical reserves of participating business (for unit-linked business the
factor is reduced to 1\%) plus a factor of 0.3\% to the sum-at-risk" (CEA and
Mercer Oliver Wyman, \textit{Solvency Assessment Models Compared, }http://www.cea.assur.org/cea/download/publ/article221.pdf).

The suitable formulae should take into account the specificities of life,
non-life and reinsurance business. Various methods are, therefore, available.
To give an example, according to the French legislation (Code des Assurances,
R334-13) for the life insurance, the solvency margin (to be replaced by the
Solvency Capital Requirement for Solvency II) should be superior to the result
obtained by multiplying 0,3\% of the capital under risk with the ratio between
the capital under risk after reinsurance and the capital under risk before
reinsurance computed for the previous exercise. The latter ratio cannot be
inferior to 50\%. To keep it simple, at time $t$ the result obtained by
multiplying a constant $\zeta_{0}$ (depending on previous experience and the
type of insurance business) by the average claim per contract and by the
number of contracts $n_{t}$ should not exceed the fortune of the insurance
company:
\begin{equation}
\zeta_{0}\times n_{t}\times\text{average claim}\leq\text{fortune at time }t.
\label{solv}%
\end{equation}

Corroborating these elements, it appears obvious that the simple collective
risk model should be improved to a model involving several contracts. We
emphasize the fact that only quantitative requirements are taken into
consideration (therefore, the model covers only part of Solvency II Pillar 1 requirements).

\section{The insurance problem with several contracts}

We introduce a complete probability space $(\Omega,\mathcal{F},P).$ In order
to model the claims, as for Mnif, Sulem (2005), we use a compound Poisson
process given by a random measure $\mu(dtdy)$ on $%
\mathbb{R}
_{+}\times B,$ with $B\subset%
\mathbb{R}
_{+}\setminus\left\{  0\right\}  .$ Moreover, we assume that the compensator
of $\mu$ takes the form $dt\pi(dy)$ and that the measure $\pi$ is finite
$\pi(dy)=\beta G(dy)$ for some probability measure $G(dy)$ on $B$ and some
positive constant $\beta.$

Throughout the section, we let $Y$ denote a generic random variable
distributed according to $G(dy).$

We consider the natural filtration $(\mathcal{F}_{t})_{t\geq0}$ generated by
the random measure $\mu.$ We call retention level any $(\mathcal{F}_{t}%
)$-adapted process $(u_{t})_{t\geq0}$ which specifies that, given a claim $y$
at time $t\geq0$, the direct insurer covers $y\wedge u_{t}$ while the
reinsurance company covers the excess of loss $(y-u_{t})^{+}.$

Since we are going to consider several insurance contracts, we introduce a
function $f$ depending both on the number of insurance contracts and on the
risk taken by the company to model the claims $f:%
\mathbb{R}
_{+}\times%
\mathbb{R}
\longrightarrow%
\mathbb{R}
_{+}.$ If the company chooses some retention level $u_{t},$ then the actual
premium rate per contract is given as in Asmussen et al. (2000), or, again, in
Mnif, Sulem (2005)%
\begin{equation}
p(u_{t})=(1+k_{1})\beta\nu-(1+k_{2})\beta E\left[  f(1,(Y-u_{t})^{+})\right]
\text{ for all }t\geq0, \label{p(u)}%
\end{equation}
where $k_{i}$ are real constants satisfying $0\leq k_{1}<k_{2}$ and
\begin{equation}
\nu=\int_{B}f(1,y)G(dy)=E[f(1,Y)]. \label{nju}%
\end{equation}
The first term in (\ref{p(u)}) is the premium received from the client, while
the second term is the quantity paid to the reinsurer.

Given the initial fortune $x\geq0$ and the retention level $u,$ if $L$ stands
for the $(\mathcal{F}_{t})-$adapted process representing the cumulative
dividends paid up to the time $t,$ $n_{t}$ denotes the number of contracts of
the insurance company at time $t,$ and $X_{t}^{x,u,L}$ the fortune of the
company , then we have%
\begin{equation}
X_{t}^{x,u,L}=x+\int_{0}^{t}n_{s}p(u_{s})ds-\int_{0}^{t+}\int_{B}%
f(n_{s},y\wedge u_{s})\mu(dsdy)-\int_{0}^{t}dL_{s}. \label{edsn}%
\end{equation}
If we denote by $a$ the quantity%
\[
a=\frac{1}{\zeta_{0}\nu},
\]
then, from (\ref{solv}) we get that the maximum number of insurance contracts
is $n_{t}^{\max}=aX_{t}^{x,u,L}$ $.$ We have the following equation%
\begin{equation}
X_{t}^{x,u,L}=x+a\int_{0}^{t}X_{s}^{x,u,L}p(u_{s})ds-\int_{0}^{t+}\int
_{B}f(aX_{s-}^{x,u,L},y\wedge u_{s})\mu(dsdy)-\int_{0}^{t}dL_{s}, \label{SDE}%
\end{equation}
and introduce the cost functional%
\begin{equation}
J(x,u,L)=E\left[  \int_{0}^{\tau}e^{-rs}dL_{s}\right]  , \label{J}%
\end{equation}
where $r$ is some discount factor and $\tau$ is the ruin time%
\[
\tau=\inf\left\{  t\geq0:X_{t}^{x,u,L}\leq0\right\}  .
\]
Our value function $V$ will be defined as the maximum over some family of
admissible couples $(u,L)$ of the cost functional $J.$

In practice, whenever the solvency condition is not satisfied, one of the
following two events may occur. In the first case, a capital infusion from the
shareholders intervenes. In the second one, an external referee solves the
problem: either by transferring some of the contracts to other insurance
companies, or by dissolving the contracts in final phase. The Solvency II
framework states that as soon as the Solvency Capital Requirement (SCR) is not
satisfied, supervisory action will be triggered. However, if the Minimum
Capital Requirement (MCR) is not satisfied, the control authority can invoke
severe measures (including closure of the company). From the mathematical
point of view, we do not allow capital infusions, these being obtained by
taking a larger initial reserve. On the contrary, the latter events may appear
and they allow the variation of the number of contracts.

Let us now return to the function $f$ modelling the claims. It is natural to
suppose that the claims increase with the number of contracts and are null if
the company has no contract. Moreover, the claims should increase with the
risks covered and should be $0$ if dealing with no risk. If the number of
contracts is positive and the risk covered by these contracts is not null,
then the claims are expected to be strictly positive. An utility function is
usually supposed to be concave. If we are given a concave function $v$ such
that $v(0)=0,$ then
\[
v(\lambda x)\geq\lambda v(x),
\]
for any $\lambda\leq1.$ Since any nonlinearity in (\ref{SDE}) may only come
from $f,$ in order to obtain the previous property for our utility function
$V$, one should assume that $f$ is convex in the first variable. These
assumptions give

\begin{assum}
\textbf{(A1) }Suppose that the function $f:%
\mathbb{R}
_{+}\times%
\mathbb{R}
\longrightarrow%
\mathbb{R}
_{+}$ satisfies:

- $f(\cdot,y)$ is convex, non decreasing and $f(0,y)=0$ for all $y\in%
\mathbb{R}
_{+}$;

- $f(x,\cdot)$ is increasing and $f(x,0)=0$;

- $f(x,y)>0$ if $x>0$ and $y>0$;

- $f$ is uniformly continuous on $%
\mathbb{R}
_{+}\times%
\mathbb{R}
$;

- $f(x,y)$ is Lipschitz in $x$, uniformly in $y\in%
\mathbb{R}
_{+}$.
\end{assum}

One expects to cover expenditures through the premium received
\[
p(u_{t})\geq\beta E[f(1,Y\wedge u_{t})].
\]
Recall that \ $p(0)-\beta E[f(1,0)]<0$ and that $lim_{u\rightarrow\infty
}\left(  p(u)-\beta E[f(1,Y\wedge u)]\right)  >0$ (recall the definitions
(\ref{p(u)}) and (\ref{nju}) of $p$ and $\nu$, respectively) and we obtain the
existence of some $\underline{u}>0$ such that
\begin{equation}
p(u)\geq\beta E[f(1,Y\wedge u)], \label{ineg0}%
\end{equation}
for all $u\geq\underline{u}$. Thus, we are going to consider only the
retention levels $u_{t}$ satisfying
\begin{equation}
u_{t}\geq\underline{u}. \label{ineq1}%
\end{equation}

One should impose that the dividends paid at some time $t$ do not exceed the
reserve at the same time. Therefore, we call admissible strategy the couple of
$(\mathcal{F}_{t})-$adapted processes $(u,L)$ such that $u$ satisfies
(\ref{ineq1}) and $L$ is c\`{a}dl\`{a}g, non decreasing, $L_{0-}=0$ and
$L_{t}-L_{t-}\leq X_{t-}^{x,u,L}$ for almost every $(t,\omega).$ We should
first prove the existence of such admissible strategies.

\begin{rem}
If $l$ is an $(\mathcal{F}_{t})-$adapted process which is c\`{a}dl\`{a}g, non
decreasing, $l_{0-}=0,$ then, for any initial condition $x\geq0,$ and any
$(\mathcal{F}_{t})-$adapted processes $u$ which satisfies (\ref{ineq1}), there
exists a unique $\mathcal{F}_{t}-$adapted right-continuous process
$X_{t}^{x,\alpha,l}$ with left-hand limits which satisfies the equation
\begin{equation}
X_{t}^{x,u,l}=x+a\int_{0}^{t}X_{s}^{x,u,l}p(u_{s})ds-\int_{0}^{t+}\int
_{B}f(aX_{s-}^{x,u,l},y\wedge u_{s})\mu(dsdy)-\int_{0}^{t}dl_{s} \label{SDE0}%
\end{equation}
(see also Ikeda, Watanabe (1989) IV, Theorem 9.1). We define the ruin time
$\tau=\inf\left\{  t\geq0:X_{t}^{x,u,l}\leq0\right\}  $. Obviously, on
$\{t<\tau\}$ we have $\triangle l_{t}=l_{t}-l_{t-}\leq X_{t-}^{x,u,l}.$ Let us
define the process%
\[
L_{t}=l_{t}1_{\{t<\tau\}}+\left(  \triangle l_{t}\wedge X_{t-}^{x,u,l}\right)
1_{\{t=\tau\}}.
\]
We get an $(\mathcal{F}_{t})-$adapted process which is c\`{a}dl\`{a}g, non
decreasing, and $L_{0-}=0.$ Let $X^{x,u,L}$ denote the solution of
(\ref{SDE0}) with $L$ instead of $l.$ We notice that $(u,L)$ is admissible in
the sense that $L_{t}-L_{t-}\leq X_{t-}^{x,u,L}$ for almost every
$(t,\omega).$
\end{rem}

For all initial reserve $x\geq0,$ we denote by $\mathcal{A}(x)$ the set of
admissible strategies described above. The value function is defined by%
\[
V(x)=\sup_{(u,L)\in\mathcal{A}(x)}J(x,u,L).
\]

\begin{prop}
(Comparison for solutions of (\ref{SDE0})) Given two $(\mathcal{F}_{t}%
)-$adapted processes $u$ and $l$ such that $u$ satisfies (\ref{ineq1}) and $l$
is c\`{a}dl\`{a}g, non decreasing, and $l_{0-}=0,$ and two initial states
$0\leq x\leq x^{\prime},$ the solutions of (\ref{SDE0}) $X^{x,u,l}$ and
$X^{x^{\prime},u,l}$ starting from $x$ (respectively $x^{\prime})$ and
associated with the pair $(u,l)$ satisfy%
\[
X_{t}^{x,u,l}\leq X_{t}^{x^{\prime},u,l}\text{, for all }t,\text{
}P-a.s.\text{ }%
\]

\end{prop}

\begin{pf}
Let us consider the sequence of functions $\phi_{n}\in C^{1}(%
\mathbb{R}
)$ such that $\phi_{n}(x^{\prime\prime})=0$ for all $x^{\prime\prime}\leq0$,
$0\leq\phi_{n}^{\prime}(x^{\prime\prime})\leq1,$ for all $x^{\prime\prime}\in%
\mathbb{R}
$, and $\phi_{n}(x^{\prime\prime})\uparrow(x^{\prime\prime})^{+}$ as
$n\rightarrow\infty.$ A simple application of It\^{o}'s formula yields%
\begin{equation}
\phi_{n}\left(  X_{t}^{x,u,l}-X_{t}^{x^{\prime},u,l}\right)  =I_{1}+I_{2},
\label{fin}%
\end{equation}
where%
\begin{align*}
&  I_{1}=\int_{0}^{t}ap(u_{s})\left(  X_{s}^{x,u,l}-X_{s}^{x^{\prime}%
,u,l}\right)  \phi_{n}^{\prime}\left(  X_{s}^{x,u,l}-X_{s}^{x^{\prime}%
,u,l}\right)  ds,\\
&  I_{2}=\\
&  \int_{0}^{t+}\int_{B}\phi_{n}\left(  X_{s-}^{x,u,l}-X_{s-}^{x^{\prime}%
,u,l}-f\left(  aX_{s-}^{x,u,l},y\wedge u_{s}\right)  +f\left(  aX_{s-}%
^{x^{\prime},u,l},y\wedge u_{s}\right)  \right)  \mu(dsdy)\\
&  -\int_{0}^{t+}\int_{B}\phi_{n}\left(  X_{s-}^{x,u,l}-X_{s-}^{x^{\prime
},u,l}\right)  \mu(dsdy).
\end{align*}
It is obvious that
\[
I_{1}\leq C\int_{0}^{t}\left(  X_{s}^{x,u,l}-X_{s}^{x^{\prime},u,l}\right)
^{+}ds,
\]
where $C$ is a constant independent of $x$ and $x^{\prime}.$ Since $a$ can be
chosen arbitrarily small (for that, it is enough to recall $a=\frac{1}%
{\zeta_{0}\nu}$ and then choose an arbitrarily small monetary unit such that
the quantity $\nu$ becomes large), we may assume that $aK_{0}\leq1$ (here
$K_{0}$ denotes the Lipschitz constant for $f).$ Then the function
$x\longmapsto x-f(ax,y)$ is increasing for all $y\in%
\mathbb{R}
_{+}$. Therefore, we get
\[
I_{2}\leq0.
\]
Combining the two estimates for $I_{1}$ and $I_{2},$ we have%
\[
E\left[  \phi_{n}\left(  X_{t}^{x,u,l}-X_{t}^{x^{\prime},u,l}\right)  \right]
\leq C\int_{0}^{t}E\left[  \left(  X_{s}^{x,u,l}-X_{s}^{x^{\prime}%
,u,l}\right)  ^{+}\right]  ds.
\]
We allow $n\rightarrow\infty$ to obtain%
\[
E\left[  \left(  X_{t}^{x,u,l}-X_{t}^{x^{\prime},u,l}\right)  ^{+}\right]
\leq C\int_{0}^{t}E\left[  \left(  X_{s}^{x,u,l}-X_{s}^{x^{\prime}%
,u,l}\right)  ^{+}\right]  ds.
\]
Finally, Gronwall's inequality yields%
\[
E\left[  \left(  X_{t}^{x,u,l}-X_{t}^{x^{\prime},u,l}\right)  ^{+}\right]
=0.
\]
The proof of our Proposition is complete.
\end{pf}

If the initial fortune is fixed, then the company has to make a choice over
some family of admissible strategies. One may naturally wonder whether the
same strategies are valid when dealing with a greater initial reserve or not.
The answer is affirmative as proven by the following Proposition.

\begin{prop}
If $0\leq x\leq x^{\prime}$ are two initial capitals and if $(u,L)$ is an
admissible strategy for $x$, then $(u,L)$ is also admissible for $x^{\prime} $.
\end{prop}

\begin{pf}
Indeed, \ if $X_{t}^{x,u,L}$ (respectively $X_{t}^{x^{\prime},u,L})$ denote
the solutions of (\ref{SDE0}) starting from $x$ (respectively $x^{\prime})$
associated with the control pair $(u,L)$, then the comparison result yields%
\[
X_{t}^{x,u,L}\leq X_{t}^{x^{\prime},u,L},dtdP-a.e.\text{ on }[0,\infty
)\times\Omega.
\]
Now, since $L$ is admissible for $x,$ we have%
\[
L_{t}-L_{t-}\leq X_{t-}^{x,u,L}\leq X_{t-}^{x^{\prime},u,L},dtdP-a.e.,
\]
and $L$ is again admissible for $x^{\prime}.$ Moreover, if $\tau$ denotes the
ruin time for $X_{t}^{x,u,L}$ and $\tau^{\prime}$ denotes the ruin time for
$X_{t}^{x^{\prime},u,L},$ then, obviously%
\[
\tau\leq\tau^{\prime},\text{ }P-a.s.
\]

\end{pf}

As one expects, using the previous results, we find that the utility function
of the insurance company increases with the initial reserve. Since our
strategy involves a dynamic programming approach, we would like to have finite
value function. We suppose that the following assumption holds true

\begin{assum}
\textbf{(A2)} The discount factor $r$ in (\ref{J}) satisfies%
\[
r>\frac{2(1+k_{1})\beta}{\zeta_{0}}%
\]
Given an economic framework in which the discount factor $r$ is fixed, the
above assumption says that the time between two claims is great enough to
justify the demand for small solvency\ translated in the small constant
$\zeta_{0}$.
\end{assum}

Under this Assumption, we provide an upper bound estimate as well as Lipschitz
regularity of the value function.

\begin{prop}
The value function $V$ is non decreasing, enjoys the Lipschitz property and
satisfies%
\begin{equation}
V(x)\leq Kx, \label{Vbound}%
\end{equation}
for some large enough positive constant $K.$
\end{prop}

\begin{pf}
The first assertion is straightforward from the previous Proposition. In order
to establish the upper bound (\ref{Vbound}), we notice that%
\[
X_{t}^{x,u,L}\leq x+\frac{(1+k_{1})\beta}{\zeta_{0}}\int_{0}^{t}X_{s}%
^{x,u,L}ds,
\]
for all $t\geq0.$ Gronwall's inequality yields%
\begin{equation}
X_{t}^{x,u,L}\leq xe^{\frac{(1+k_{1})\beta}{\zeta_{0}}t}. \label{Xt}%
\end{equation}
We write It\^{o}'s formula for $e^{-rt}X_{t}^{x,u,L}$ and use (\ref{Xt})
together with \textbf{(A2)} to obtain%
\[
J(x,u,L)\leq Cx.
\]
Here $C$ is a constant which may change from line to line. Let us fix
$x,x^{\prime}\geq0.$ Suppose that $(u,L)\in\mathcal{A}(x+x^{\prime})$ and
notice that, in this case, $\left(  u,\frac{x}{x+x^{\prime}}L\right)
\in\mathcal{A}(x).$ Indeed,%
\begin{align*}
X_{t}^{x+x^{\prime},u,L}  &  =(x+x^{\prime})+a\int_{0}^{t}X_{s}^{x+x^{\prime
},u,L}p(u_{s})ds\\
&  -\int_{0}^{t+}\int_{B}f(aX_{s-}^{x+x^{\prime},u,L},y\wedge u_{s}%
)\mu(dsdy)-\int_{0}^{t}dL_{s},
\end{align*}
and, by multiplying the latter equality by $\frac{x}{x+x^{\prime}}$, we get%
\begin{align*}
\frac{x}{x+x^{\prime}}X_{t}^{x+x^{\prime},u,L}  &  =x+\int_{0}^{t}%
ap(u_{s})\frac{x}{x+x^{\prime}}X_{s}^{x+x^{\prime},u,L}ds\\
&  -\int_{0}^{t+}\int_{B}\frac{x}{x+x^{\prime}}f\left(  aX_{s-}^{x+x^{\prime
},u,L},y\wedge u_{s}\right)  \mu(dsdy)\\
&  -\int_{0}^{t}d\left(  \frac{x}{x+x^{\prime}}L_{s}\right)  .
\end{align*}
On the other hand,%
\begin{align*}
X_{t}^{x,u,\frac{x}{x+x^{\prime}}L}  &  =x+\int_{0}^{t}ap(u_{s})X_{s}%
^{x,u,\frac{x}{x+x^{\prime}}L}ds\\
&  -\int_{0}^{t+}\int_{B}f\left(  aX_{s-}^{x^{\prime},u,\frac{x}{x+x^{\prime}%
}L},y\wedge u_{s}\right)  \mu(dsdy)\\
&  -\int_{0}^{t}d\left(  \frac{x}{x+x^{\prime}}L_{s}\right)  .
\end{align*}
Now, let the functions $\phi_{n}\in C^{1}(%
\mathbb{R}
)$ be such that $\phi_{n}(x^{\prime\prime})=0$ for all $x^{\prime\prime}\leq
0$, and $0\leq\phi_{n}^{\prime}(x^{\prime\prime})\leq1,$ for all
$x^{\prime\prime}\in%
\mathbb{R}
$, and $\phi_{n}(x^{\prime\prime})\uparrow(x^{\prime\prime})^{+}$ as
$n\rightarrow\infty.$ We make the following notation%
\[
\frac{x}{x+x^{\prime}}X_{\cdot}^{x+x^{\prime},u,L}=Y_{\cdot}.
\]
We apply It\^{o}'s formula to have%
\begin{equation}
\phi_{n}\left(  Y_{t}-X_{t}^{x,u,\frac{x}{x+x^{\prime}}L}\right)  =I_{1}%
+I_{2}, \label{fin2}%
\end{equation}
where%
\begin{align*}
&  I_{1}=\int_{0}^{t}ap(u_{s})\left(  Y_{s}-X_{s}^{x,u,\frac{x}{x+x^{\prime}%
}L}\right)  \phi_{n}^{\prime}\left(  Y_{s}-X_{s}^{x,u,\frac{x}{x+x^{\prime}}%
L}\right)  ds,\\
&  I_{2}=\\
&  \int_{0}^{t+}\int_{B}\phi_{n}\left(  Y_{s-}-X_{s-}^{x,u,\frac
{x}{x+x^{\prime}}L}-\frac{x}{x+x^{\prime}}f\left(  aX_{s-}^{x+x^{\prime}%
,u,L},y\wedge u_{s}\right)  +f\left(  aX_{s-}^{x,u,\frac{x}{x+x^{\prime}}%
L},y\wedge u_{s}\right)  \right)  \mu(dsdy)\\
&  -\int_{0}^{t+}\int_{B}\phi_{n}\left(  Y_{s-}-X_{s-}^{x,u,\frac
{x}{x+x^{\prime}}L}\right)  \mu(dsdy).
\end{align*}
It is obvious that
\[
I_{1}\leq C\int_{0}^{t}\left(  Y_{s}-X_{s}^{x,u,\frac{x}{x+x^{\prime}}%
L}\right)  ^{+}ds,
\]
where $C$ is a constant independent of $x$ and $x^{\prime}$, and we use the
convexity of $f$ in the first variable and $f(0,\cdot)=0$, together with the
monotonicity of $\phi_{n}$ to get (as in the proof of the comparison result),
\[
I_{2}\leq0.
\]
Thus we obtain, as in the comparison result,%
\[
\frac{x}{x+x^{\prime}}X_{t}^{x+x^{\prime},u,L}\leq X_{t}^{x,u,\frac
{x}{x+x^{\prime}}L}\text{ }dtdP-a.e.\text{ on }[0,\infty)\times\Omega.
\]
Obviously, $\left(  u,\frac{x}{x+x^{\prime}}L\right)  $ is an admissible
strategy for the initial reserve $x.$ If $\tau$ is the ruin time for the
strategy $(u,L)$ for the initial reserve $x+x^{\prime},$ then the above
inequality states that the ruin time for the strategy $\left(  u,\frac
{x}{x+x^{\prime}}L\right)  $ when the initial reserve is $x$ is greater than
or equal to $\tau.$ Therefore, we have%
\[
V(x+x^{\prime})=\frac{x+x^{\prime}}{x}\sup_{(u,L)\in\mathcal{A}(x+x^{\prime}%
)}E\left[  \int_{0}^{\tau}e^{-rs}d\left(  \frac{x}{x+x^{\prime}}L_{s}\right)
\right]  \leq\frac{x+x^{\prime}}{x}V(x).
\]
and (\ref{Vbound}) gives the Lipschitz property of $V.$ The proof of the
Proposition is complete.
\end{pf}

\section{Hamilton Jacobi Bellman Variational Inequality}

We have already seen that our value function $V$ is increasing and Lipschitz
continuous. These properties allow us to prove in a standard way that $V$
satisfies the following \textbf{Dynamic Programming Principle}

\begin{prin}
\textbf{(DPP)}%
\[
V(x)=\sup_{(u,L)\in\mathcal{A}(x)}E\left[  e^{-r(t\wedge\tau)}V(X_{t\wedge
\tau}^{x,u,L})+\int_{0}^{t\wedge\tau}e^{-rs}dL_{s}\right]  ,
\]
for all $t\geq0,$ $x\geq0.$
\end{prin}

For further literature on the subject, the reader to referred to Fleming,
Soner (1993), Krylov (1980), or Yong, Zhou (1999) (theorem 4.3.3), for
diffusion state processes or to Pham (1998) in the case of jump diffusion processes.

We consider at this point the following HJB variational inequality:%
\begin{equation}
\left\{
\begin{array}
[c]{l}%
\max\{H(x,V,V^{\prime}(x)),1-V^{\prime}(x)\}=0\text{ in }%
\mathbb{R}
_{+}^{\ast},\\
V(0)=0.
\end{array}
\right.  , \label{HJBVI}%
\end{equation}
where%
\begin{align}
&  H(x,V,q)\label{H}\\
&  =\sup_{u\geq\underline{u}}\left\{  -rV(x)+axp(u)q+\int_{B}\left[
V(x-f(ax,y\wedge u))-V(x)\right]  \pi(dy)\right\}  .\nonumber
\end{align}

Let us recall that $C^{1,1}(%
\mathbb{R}
_{+})$ stands for the class of all real-valued, differentiable functions on $%
\mathbb{R}
_{+}$ such that the derivative is locally Lipschitz.

We also recall the definition of the viscosity supersolution, respectively
viscosity subsolution.

\begin{defn}
(i) Any lower semi-continuous (respectively upper semi-continuous) function
$v$ is a viscosity supersolution (subsolution) of (\ref{HJBVI}) if $v(0)\geq0
$ ($\leq0$) and%
\[
\max\left\{  H(x,\varphi,\varphi^{\prime}(x)),1-\varphi^{\prime}(x)\right\}
\leq0,
\]
(respectively $\ \geq0)$ whenever $\varphi\in C^{1,1}(%
\mathbb{R}
_{+})$ is such that $v-\varphi$ has a global minimum (maximum) at $x>0$.

(ii) A function $v$ is a viscosity solution of (\ref{HJBVI}) if it is both
super and subsolution.
\end{defn}

\begin{thm}
The value function $V$ is a viscosity solution for the associated
Hamilton-Jacobi-Bellman Variational Inequality\ (\ref{HJBVI}).
\end{thm}

\begin{pf}
First, we prove that $V$ is a viscosity supersolution for (\ref{HJBVI}). In
order to do this, let us consider $x\in%
\mathbb{R}
_{+}^{\ast}$ and a $C^{1,1}$ test function $\varphi$ such that $V(x^{\prime
})-\varphi(x^{\prime})\geq V(x)-\varphi(x)=0$, for all $x^{\prime}\in%
\mathbb{R}
_{+}^{\ast}.$ Moreover, consider $0<h<x$ and the admissible strategy
$(u,L)\in\mathcal{A}(x)$ where $L_{s}=h$, for all $s\geq0$ and $u$ is
admissible and arbitrarily chosen. We have%
\begin{align*}
\varphi(x)  &  =V(x)\geq E\left[  \int_{0}^{t\wedge\tau}e^{-rs}dL_{s}%
+e^{-r(t\wedge\tau)}V(X_{t\wedge\tau}^{x,u,L})\right] \\
&  \geq h+E\left[  e^{-r(t\wedge\tau)}\varphi(X_{t\wedge\tau}^{x,u,L})\right]
,
\end{align*}
for all $t\geq0.$ We take the limit as $t\rightarrow0+$ and get%
\[
\varphi(x)\geq h+\varphi(x-h).
\]
This latter inequality yields%
\begin{equation}
1-\varphi^{\prime}(x)\leq0. \label{superin2}%
\end{equation}

In order to prove $H(x,\varphi,\varphi^{\prime}(x))\leq0,$ we consider the
admissible pair $L_{s}=0,$ $u_{s}=u_{0},$ for all $s\geq0$ (here $u_{0}%
\geq\underline{u}$ is arbitrarily chosen). We apply It\^{o}'s formula to
$e^{-r(t\wedge\tau)}\varphi(X_{t\wedge\tau}^{x,u,L})$ to obtain%
\begin{align*}
&  E\left[  e^{-r(t\wedge\tau)}\varphi(X_{t\wedge\tau}^{x,u,L})\right]
-\varphi(x)\\
&  =E\left[  \int_{0}^{t\wedge\tau}\left(  -re^{-rs}\varphi(X_{s}%
^{x,u,L})+e^{-rs}aX_{s}^{x,u,L}p(u_{0})\varphi^{\prime}(X_{s}^{x,u,L})\right)
ds\right] \\
&  +E\left[  \int_{0}^{t\wedge\tau}\int_{B}e^{-rs}\left(  \varphi\left(
X_{s-}^{x,u,L}-f\left(  aX_{s-}^{x,u,L},y\wedge u_{0}\right)  \right)
-\varphi\left(  X_{s-}^{x,u,L}\right)  \right)  \mu(dsdy)\right]  .
\end{align*}
Recalling that $\varphi(x)\geq E\left[  e^{-r(t\wedge\tau)}V(X_{t\wedge\tau
}^{x,u,L})\right]  ,$ and dividing by $t>0,$ we have
\begin{align}
0  &  \geq E\left[  \frac{1}{t}\int_{0}^{t\wedge\tau}\left(  -re^{-rs}%
\varphi(X_{s}^{x,u,L})+e^{-rs}aX_{s}^{x,u,L}p(u_{0})\varphi^{\prime}%
(X_{s}^{x,u,L})\right)  ds\right] \nonumber\\
&  +E\left[  \frac{1}{t}\int_{0}^{t\wedge\tau}\int_{B}e^{-rs}\left(
\varphi\left(  X_{s-}^{x,u,L}-f\left(  aX_{s-}^{x,u,L},y\wedge u_{0}\right)
\right)  -\varphi\left(  X_{s-}^{x,u,L}\right)  \right)  \mu(dsdy)\right]
\nonumber\\
&  \geq E\left[  \frac{1}{t}\int_{0}^{t\wedge\tau}\left(  -r\varphi
(x)+e^{-rt}axp(u_{0})\varphi^{\prime}(x)\right)  ds\right] \nonumber\\
&  +E\left[  \frac{1}{t}\int_{0}^{t\wedge\tau}ds\int_{B}\left(  e^{-rt}%
\varphi\left(  x-f\left(  ax,y\wedge u_{0}\right)  \right)  -\varphi\left(
x\right)  \right)  \pi(dy)\right] \nonumber\\
&  -O\left(  E\left[  \sup_{s\leq t\wedge\tau}e^{-rs}\left\vert X_{s}%
^{x,u,L}-x\right\vert \right]  \right)  , \label{superin1.1}%
\end{align}
where $O(\delta)\rightarrow0$ whenever $\delta\rightarrow0.$

We wish to prove that $E\left[  \sup_{s\leq t\wedge\tau}e^{-rs}\left\vert
X_{s}^{x,u,L}-x\right\vert \right]  \rightarrow0$, when $t\rightarrow0.$ In
order to do this, we use%
\begin{align*}
\left\vert X_{s}^{x,u,L}-x\right\vert  &  \leq\int_{0}^{s}ap(u_{0}%
)X_{s^{\prime}}^{x,u,L}ds^{\prime}\\
&  +\int_{0}^{s+}\int_{B}f(aX_{s^{\prime}-}^{x,u,L},y\wedge u_{0}%
)\mu(ds^{\prime}dy).
\end{align*}
Therefore, with the notation $C_{0}=\frac{(1+k_{1})\beta}{\zeta_{0}}$, we
have, for some constant $C,$%
\begin{align*}
\left\vert X_{s}^{x,u,L}-x\right\vert  &  \leq x\left(  e^{C_{0}s}-1\right) \\
&  +Cx\int_{0}^{s+}\int_{B}e^{C_{0}s^{\prime}}\mu(ds^{\prime}dy)
\end{align*}
for all $0\leq s\leq t\wedge\tau$ (we use the Lipschitz property of $f$ in $x$
uniformly in $y,$ $f(0,\cdot)=0$ and the upper bound for $X_{s^{\prime}%
}^{x,u,L}$ given by (\ref{Xt})) . We multiply the last inequality by
$e^{-rs},$ take the supremum over all $0\leq s\leq t\wedge\tau$, then the
expectation with respect to $P$ to obtain%
\begin{equation}
\lim_{t\rightarrow0+}E\left[  \sup_{s\leq t\wedge\tau}e^{-rs}\left\vert
X_{s}^{x,u,L}-x\right\vert \right]  =0. \label{ineq2}%
\end{equation}
Notice that%
\[
\frac{E[t\wedge\tau]}{t}\geq1-P(\tau\leq t)\geq1-P(\eta_{1}\leq t),
\]
where $\eta_{1}$ is the first time a claim occurs (it follows the exponential
law). Consequently,%
\begin{equation}
\lim_{t\rightarrow0+}\frac{E[t\wedge\tau]}{t}=1. \label{ineq3}%
\end{equation}
Returning to (\ref{superin1.1}) we let $t\rightarrow0+$ and use (\ref{ineq2})
and (\ref{ineq3}) to get%
\begin{align}
0  &  \geq\left(  -r\varphi(x)+axp(u_{0})\varphi^{\prime}(x)\right)
\label{superin1}\\
&  +\int_{B}\left\{  \varphi\left(  x-f\left(  ax,y\wedge u_{0}\right)
\right)  -\varphi\left(  x\right)  \right\}  \pi(dy)\nonumber
\end{align}
Combining (\ref{superin1}) and (\ref{superin2}), we prove that $V$ is a
viscosity supersolution for (\ref{HJBVI}).

In order to prove that the value function is a viscosity subsolution for
(\ref{HJBVI}), we fix $x>0$ and consider an arbitrary test function
$\varphi\in C^{1,1}$ such that $V(x^{\prime})-\varphi(x^{\prime})\leq
V(x)-\varphi(x)=0,$ for all $x^{\prime}\in%
\mathbb{R}
_{+}.$ Let us suppose that the subsolution inequality does not hold.
Therefore, there exists $\delta>0$ such that%
\[
\max\left\{  H(x,\varphi,\varphi^{\prime}(x)),1-\varphi^{\prime}(x)\right\}
<-\delta.
\]
We use the continuity of $H$ and of $\varphi^{\prime}$ to obtain the existence
of some $\eta\in\left(  0,x\wedge\frac{\delta}{4K_{\varphi}}\right)  ,$ where
$K_{\varphi}$ denotes the Lipschitz constant for $\varphi$ on $\left[
0,e^{r}x\right]  ,$ such that
\begin{equation}
\max\left\{  H(x^{\prime},\varphi,\varphi^{\prime}(x^{\prime})),1-\varphi
^{\prime}(x^{\prime})\right\}  <-\delta,\text{ }if\text{ }x^{\prime}\in
B(x,\eta). \label{star}%
\end{equation}

Let us consider an arbitrary strategy $(u,L)\in\mathcal{A}(x)$ and let
$X^{x,u,L}$ denote the solution of (\ref{SDE0}) for $(u,L)$ instead of $(u,l).
$ We define the stopping time%
\[
\sigma=\inf\{t\geq0:\text{ }X_{t}^{x,u,L}\notin B(x,\eta)\}.
\]
Obviously $\sigma\leq\tau$ (the ruin time). We apply It\^{o}'s formula to
$e^{-r(t\wedge\sigma)}\varphi(X_{t\wedge\sigma}^{x,u,L})$ and write%
\begin{align}
&  E\left[  e^{-r(t\wedge\sigma)}\varphi(X_{t\wedge\sigma}^{x,u,L})\right]
-\varphi(x)=\label{subin1}\\
&  E\left[  \int_{0}^{t\wedge\sigma}\left(  -re^{-rs}\varphi(X_{s}%
^{x,u,L})+e^{-rs}aX_{s}^{x,u,L}p(u_{0})\varphi^{\prime}(X_{s}^{x,u,L})\right)
ds\right] \nonumber\\
&  +E\left[  \int_{0}^{t\wedge\sigma}\int_{B}e^{-rs}\left(  \varphi\left(
X_{s-}^{x,u,L}-f\left(  aX_{s-}^{x,u,L},y\wedge u_{0}\right)  \right)
-\varphi\left(  X_{s-}^{x,u,L}\right)  \right)  \mu(dsdy)\right] \nonumber\\
&  -E\left[  \int_{0}^{t\wedge\sigma}e^{-rs}\varphi^{\prime}\left(
X_{s}^{x,u,L}\right)  dL_{s}^{c}\right] \nonumber\\
&  +E\left[  \sum_{s\leq t\wedge\sigma}e^{-rs}\left(  \varphi\left(
X_{s-}^{x,u,L}-\triangle L_{s}\right)  -\varphi\left(  X_{s-}^{x,u,L}\right)
\right)  \right] \nonumber
\end{align}
For $s<t\wedge\sigma$ we have, from (\ref{star})%
\begin{align}
-  &  re^{-rs}\varphi(X_{s}^{x,u,L})+e^{-rs}aX_{s}^{x,u,L}p(u_{0}%
)\varphi^{\prime}(X_{s}^{x,u,L})\label{subin2}\\
+  &  e^{-rs}\int_{B}\left(  \varphi\left(  X_{s}^{x,u,L}-f\left(
aX_{s}^{x,u,L},y\wedge u_{0}\right)  \right)  -\varphi\left(  X_{s}%
^{x,u,L}\right)  \right)  \pi(dy)<-\delta e^{-rs},\nonumber
\end{align}
and, again from (\ref{star}),
\[
\varphi^{\prime}\left(  X_{s}^{x,u,L}\right)  >1.
\]
It follows that%
\begin{equation}
\varphi\left(  X_{s-}^{x,u,L}-\triangle L_{s}\right)  -\varphi\left(
X_{s-}^{x,u,L}\right)  \leq-\triangle L_{s}. \label{subin3}%
\end{equation}
Using (\ref{subin2}) we get%
\begin{align}
&  E\left[  \int_{0}^{t\wedge\sigma}\left(  -re^{-rs}\varphi(X_{s}%
^{x,u,L})+e^{-rs}aX_{s}^{x,u,L}p(u_{0})\varphi^{\prime}(X_{s}^{x,u,L})\right)
ds\right] \nonumber\\
+  &  E\left[  \int_{0}^{t\wedge\sigma}\int_{B}e^{-rs}\left(  \varphi\left(
X_{s-}^{x,u,L}-f\left(  aX_{s-}^{x,u,L},y\wedge u_{0}\right)  \right)
-\varphi\left(  X_{s-}^{x,u,L}\right)  \right)  \mu(dsdy)\right] \nonumber\\
\leq &  \delta E\left[  \frac{e^{-r(t\wedge\sigma)}-1}{r}\right] \nonumber\\
-  &  E\left[  \int_{0}^{t\wedge\sigma}\int_{B}e^{-rs}\left(  \varphi\left(
X_{s}^{x,u,L}-f\left(  aX_{s}^{x,u,L},y\wedge u_{0}\right)  \right)
-\varphi\left(  X_{s}^{x,u,L}\right)  \right)  \pi(dy)ds\right] \nonumber\\
+  &  \int_{0}^{t\wedge\sigma}ds\int_{B}e^{-rs}\left(  \varphi\left(
x-f\left(  ax,y\wedge u_{0}\right)  \right)  -\varphi\left(  x\right)
\right)  \pi\left(  dy\right) \nonumber\\
-  &  E\left[  \int_{0}^{t\wedge\sigma}\int_{B}e^{-rs}\left(  \varphi\left(
x-f\left(  ax,y\wedge u_{0}\right)  \right)  -\varphi\left(  x\right)
\right)  \mu(dsdy)\right] \nonumber\\
+  &  E\left[  \int_{0}^{t\wedge\sigma}\int_{B}e^{-rs}\left(  \varphi\left(
X_{s-}^{x,u,L}-f\left(  aX_{s-}^{x,u,L},y\wedge u_{0}\right)  \right)
-\varphi\left(  X_{s-}^{x,u,L}\right)  \right)  \mu(dsdy)\right] \nonumber\\
\leq &  \delta E\left[  \frac{e^{-r(t\wedge\sigma)}-1}{r}\right]
+4K_{\varphi}\eta E\left[  \frac{1-e^{-r(t\wedge\sigma)}}{r}\right]  .
\label{subin4}%
\end{align}
We return to (\ref{subin1}) and use (\ref{subin3}) and (\ref{subin4}) to get%
\begin{align*}
E\left[  e^{-r(t\wedge\sigma)}\varphi(X_{t\wedge\sigma}^{x,u,L})\right]
-\varphi(x)  &  \leq\delta E\left[  \frac{e^{-r(t\wedge\sigma)}-1}{r}\right]
+4K_{\varphi}\eta E\left[  \frac{1-e^{-r(t\wedge\sigma)}}{r}\right] \\
&  -E\left[  \int_{0}^{t\wedge\sigma}e^{-rs}dL_{s}\right]  ,
\end{align*}
and, from this,
\begin{align}
V(x)  &  =\varphi(x)\nonumber\\
&  \geq E\left[  e^{-r(t\wedge\sigma)}\varphi(X_{t\wedge\sigma}^{x,u,L}%
)+\int_{0}^{t\wedge\sigma}e^{-rs}dL_{s}\right] \nonumber\\
&  +\left(  \delta-4K_{\varphi}\eta\right)  E\left[  \frac{1-e^{-r(t\wedge
\sigma)}}{r}\right]  .\label{subin5}\\
&  \geq E\left[  e^{-r(t\wedge\sigma)}\varphi(X_{t\wedge\sigma}^{x,u,L}%
)+\int_{0}^{t\wedge\sigma}e^{-rs}dL_{s}\right]  +\frac{\delta-4K_{\varphi}%
\eta}{2}E[t\wedge\sigma],\nonumber
\end{align}
for $t$ small enough. We can suppose that $x$ is a strict global maximum
point. Then there exists $\lambda>0$ such that%
\[
\sup_{x^{\prime}\notin B%
{{}^\circ}%
(x,\eta)}(V(x^{\prime})-\varphi(x^{\prime}))=-\lambda.
\]
We use (\ref{subin5}) and write
\begin{align}
V(x)  &  \geq E\left[  e^{-r(t\wedge\sigma)}V(X_{t\wedge\sigma}^{x,u,L}%
)+\int_{0}^{t\wedge\sigma}e^{-rs}dL_{s}\right] \nonumber\\
+  &  \lambda E\left[  e^{-r(t\wedge\sigma)}1_{\sigma\leq t}\right]
+\frac{\delta-4K_{\varphi}\eta}{2}tP(\sigma>t)\nonumber\\
&  \geq E\left[  e^{-r(t\wedge\sigma)}V(X_{t\wedge\sigma}^{x,u,L})+\int
_{0}^{t\wedge\sigma}e^{-rs}dL_{s}\right] \nonumber\\
+  &  \left(  \lambda e^{-rt}\right)  \wedge\left(  \frac{\delta-4K_{\varphi
}\eta}{2}t\right)  \label{subin7}%
\end{align}
The dynamic programming principle yields
\begin{equation}
V(x)\leq\sup_{(u,L)}E\left[  e^{-r(t\wedge\sigma)}V(X_{t\wedge\sigma}%
^{x,u,L})+\int_{0}^{t\wedge\sigma}e^{-rs}dL_{s}\right]  . \label{subin6}%
\end{equation}
Therefore, by the choice of $\eta<\frac{\delta}{4K_{\varphi}}$ and
$\lambda>0,$ (\ref{subin7}) contradicts (\ref{subin6}). This proves that $V$
is a viscosity subsolution for (\ref{HJBVI}). Our Theorem is now complete.
\end{pf}

\section{The Comparison Theorem}

The following Lemma provides an equivalent definition for the notions of
viscosity super and subsolution.

\begin{lem}
(i) A continuous function $U$ is a viscosity supersolution for (\ref{HJBVI})
in $%
\mathbb{R}
_{+}^{\ast}$ if and only if, $U(0)\geq0$ and, for any $x\in%
\mathbb{R}
_{+}^{\ast}$ and any test function $\varphi\in C^{1,1}$ such that $U-\varphi$
has a global strict minimum at $x,$ we have%
\begin{equation}
\max\left\{  H(x,U,\varphi^{\prime}(x)),1-\varphi^{\prime}(x)\right\}  \leq0.
\label{HJBVI1'}%
\end{equation}

(ii) A continuous function $U$ is a viscosity subsolution for (\ref{HJBVI}) in
$%
\mathbb{R}
_{+}^{\ast}$ if and only if, $U(0)\leq0$ and, for any $x\in%
\mathbb{R}
_{+}^{\ast}$ and any test function $\varphi\in C^{1,1}$ such that $U-\varphi$
has a global strict maximum at $x,$ we have
\begin{equation}
\max\left\{  H(x,U,\varphi^{\prime}(x)),1-\varphi^{\prime}(x)\right\}  \geq0.
\label{HJBVI2'}%
\end{equation}

\end{lem}

\begin{pf}
We only prove the assertion for viscosity supersolution, the proof for
subsolution being similar.

Suppose that (i) holds true. For any test function $\varphi\in C^{1,1}$such
that $U(x)=\varphi(x)$ and $U-\varphi$ has a global minimum at $x$, and all
$\delta>0$, we define%
\[
\varphi_{\delta}(x^{\prime})=\varphi\left(  x^{\prime}\right)  -\delta
\left\vert x^{\prime}-x\right\vert ^{2},\text{ for all }x^{\prime}\in%
\mathbb{R}
_{+}^{\ast}.
\]
Then $\varphi_{\delta}\in C^{1,1}$ and $U-\varphi_{\delta}$ has a global
strict minimum at $x$. The assumption implies that
\[
\max\left\{  H(x,U,\varphi_{\delta}^{\prime}(x)),1-\varphi_{\delta}^{\prime
}(x)\right\}  \leq0.
\]
Obviously, $U(x^{\prime})-U(x)>\varphi_{\delta}(x^{\prime})-\varphi_{\delta
}(x)$, for all $x^{\prime}\in%
\mathbb{R}
_{+}^{\ast}\smallsetminus\left\{  x\right\}  $ . The definition of $H$,
together with the last inequality, yields%
\begin{equation}
\max\left\{  H(x,\varphi_{\delta},\varphi^{\prime}(x)),1-\varphi^{\prime
}(x)\right\}  \leq0. \label{delta}%
\end{equation}
Moreover, again from the definition of $H$,%
\begin{align*}
H(x,\varphi,\varphi^{\prime}(x))  &  \leq H(x,\varphi_{\delta},\varphi
^{\prime}(x))\\
&  +\sup_{u\geq\underline{u}}\left(  \int_{B}\left(  \varphi\left(
x-f(ax,y\wedge u\right)  -\varphi_{\delta}\left(  x-f(ax,y\wedge u\right)
\right)  \pi\left(  dy\right)  \right) \\
&  \leq H(x,\varphi_{\delta},\varphi^{\prime}(x))+C\delta,
\end{align*}
where $C$ is a generic constant independent of $\delta$. We get, using
(\ref{delta}) then taking the limit as $\delta\searrow0$,
\[
\max\left\{  H(x,\varphi,\varphi^{\prime}(x)),1-\varphi^{\prime}(x)\right\}
\leq0.
\]
For the converse, consider an arbitrary test function $\varphi\in C^{1,1}$ and
$x\in%
\mathbb{R}
_{+}^{\ast}$ such that
\[
0=U(x)-\varphi(x)<U(x^{\prime})-\varphi(x^{\prime}),
\]
for all $x^{\prime}\in%
\mathbb{R}
_{+}^{\ast}\setminus\{x\}.$ For $\varepsilon>0$ such that $\varepsilon
<\frac{x}{4},$ we define%
\[
\delta_{\varepsilon}=\sup_{x^{\prime}\in B\left(  x,4\varepsilon\right)
}\left(  U(x^{\prime})-\varphi(x^{\prime})\right)  >0.
\]
It is obvious that $\lim_{\varepsilon\rightarrow0}\searrow\delta_{\varepsilon
}=0.$ We introduce%
\[
\varphi_{\varepsilon}=\left(  U-\varphi-\delta_{\varepsilon}\right)  1_{
\left[  0,x-2\varepsilon\right]  }+\left(  U(0)-\varphi(0)-\delta
_{\varepsilon}\right)  1_{%
\mathbb{R}
_{-}}.
\]
We consider some sequence of mollifiers $\rho_{n}\in C_{c}^{\infty}\left(
\mathbb{R}
;%
\mathbb{R}
_{+}\right)  ,$ $Supp$ $\rho_{n}\subset B\left(  0,\frac{1}{n}\right)  $ and
$\int_{%
\mathbb{R}
}\rho_{n}(t)dt=1$. Since $U-\varphi$ is continuous, the sequence $\left\{
\rho_{n}\ast\varphi_{\varepsilon}\right\}  _{n}$ converges uniformly on
$[0,x-3\varepsilon]$ to $\varphi_{\varepsilon}$. Then there exists a
subsequence (denoted by $\left(  \rho_{\varepsilon}\right)  $) such that
$Supp$ $\rho_{\varepsilon}\subset B\left(  0,\varepsilon\right)  $ and
\[
U\left(  x^{\prime}\right)  -\varphi\left(  x^{\prime}\right)  -2\delta
_{\varepsilon}\leq\left(  \rho_{\varepsilon}\ast\varphi_{\varepsilon}\right)
(x^{\prime})<U\left(  x^{\prime}\right)  -\varphi\left(  x^{\prime}\right)  ,
\]
for all $0\leq x^{\prime}\leq x-3\varepsilon$ and all $\varepsilon>0$.
Finally, we define the function
\[
F_{\varepsilon}(x^{\prime})=\varphi\left(  x^{\prime}\right)  +\left(
\rho_{\varepsilon}\ast\varphi_{\varepsilon}\right)  \left(  x^{\prime}\right)
.
\]
It is obvious that $F_{\varepsilon}\in C^{1,1}$ has the following properties:%
\[
\left\{
\begin{array}
[c]{l}%
F_{\varepsilon}\left(  x^{\prime}\right)  =\varphi\left(  x^{\prime}\right)
,\text{ if }x^{\prime}\geq x-\varepsilon,\\
U\left(  x^{\prime}\right)  -2\delta_{\varepsilon}\leq F_{\varepsilon}\left(
x^{\prime}\right)  \text{, if }0\leq x^{\prime}\leq x-3\varepsilon,\\
F_{\varepsilon}\left(  x^{\prime}\right)  <U\left(  x^{\prime}\right)  ,\text{
if }x^{\prime}\neq x.
\end{array}
\right.
\]
The assumptions give%
\[
\max\left\{  H(x,F_{\varepsilon},F_{\varepsilon}^{\prime}(x)),1-F_{\varepsilon
}^{\prime}(x)\right\}  \leq0.
\]
Let us put
\begin{align*}
G(x^{\prime})  &  =\sup_{u\geq\underline{u}}\left\{  -r\left(  U(x^{\prime
})-F_{\varepsilon}(x^{\prime})\right)  +ap(u)x^{\prime}\left(  \varphi
^{\prime}(x^{\prime})-F_{\varepsilon}^{\prime}(x^{\prime})\right)  \right. \\
&  +\int_{B}\left(  U(x^{\prime}-f(ax^{\prime},y\wedge u))-F_{\varepsilon
}\left(  x^{\prime}-f(ax^{\prime},y\wedge u)\right)  \right)  \pi(dy)\\
&  -\int_{B}\left(  U(x^{\prime})-F_{\varepsilon}(x^{\prime})\right)
\pi(dy)\},
\end{align*}
for any $x^{\prime}\in%
\mathbb{R}
_{+}^{\ast}.$ Then%
\begin{equation}
H(x,U,\varphi^{\prime}(x))-H(x,F_{\varepsilon},F_{\varepsilon}^{\prime
}(x))\leq G(x), \label{ineq5}%
\end{equation}
where%
\begin{align*}
G(x)  &  \leq\sup_{u\geq\underline{u}}\left\{  \int_{B}\left(
U(x-f(ax,y\wedge u))-F_{\varepsilon}\left(  x-f(ax,y\wedge u)\right)  \right)
\pi(dy)\right. \\
&  -\left.  \int_{B}\left(  U(x)-F_{\varepsilon}(x)\right)  \pi(dy)\right\}  .
\end{align*}
We then consider the sets $B^{u}=\left\{  y\in B:x-f(ax,y\wedge u)\in
\overline{B}(x,3\varepsilon)\right\}  \ $and get%
\begin{align}
&  \int_{B}\left(  U(x-f(ax,y\wedge u))-F_{\varepsilon}\left(  x-f(ax,y\wedge
u)\right)  \right)  \pi(dy)\nonumber\\
&  \leq\int_{B\setminus B^{u}}2\delta_{\varepsilon}\pi(dy)+C\pi(B^{u}%
)\nonumber\\
&  \leq2\beta\delta_{\varepsilon}+C\pi(B^{u}), \label{maj}%
\end{align}
where $C>0$ is a generic constant independent of $\varepsilon$. Moreover, if
$y\in B^{u}$, then
\[
x-f(ax,y\wedge u)\geq x-3\varepsilon.
\]
Therefore,%
\[
f(ax,y\wedge\underline{u})\leq f(ax,y\wedge u)\leq3\varepsilon.
\]
Since $f(ax,y\wedge\underline{u})>0$ for $xy>0$ and $f(ax,\cdot)$ is
nondecreasing, we deduce the existence of some $\eta_{\varepsilon}>0$ such
that $\eta_{\varepsilon}\rightarrow0$ as $\varepsilon\rightarrow0$ and $y\in
B^{u}$ only if $y\leq\eta_{\varepsilon}.$ Thus, returning to (\ref{maj}), we
get%
\begin{align*}
&  \int_{B}\left(  U(x-f(ax,y\wedge u))-F_{\varepsilon}\left(  x-f(ax,y\wedge
u)\right)  \right)  \pi(dy)\\
&  \leq C\delta_{\varepsilon}+C\pi(B\cap\lbrack0,\eta_{\varepsilon}])\text{.}%
\end{align*}
Consequently,%
\begin{equation}
G(x)\leq C\delta_{\varepsilon}+C\pi(B\cap\lbrack0,\eta_{\varepsilon}])\text{.}
\label{ineq4}%
\end{equation}
Recall that $0\notin B$. Thus, using (\ref{ineq4}) in (\ref{ineq5}) and taking
the limit as $\varepsilon\rightarrow0,$ we obtain%
\[
H(x,U,\varphi^{\prime})\leq0,
\]
and (i) follows.
\end{pf}

The assertion (ii) follows in the same way.

Under the assumption \textbf{(A2) }we are able to prove the following result
on the comparison of viscosity solutions for (\ref{HJBVI}).

\begin{thm}
Let $U$ and $V$ be respectively a continuous viscosity subsolution and a
continuous viscosity supersolution for (\ref{HJBVI}) both of at most linear
growth$.$ Then, if \textbf{(A2)} holds true, we have
\[
U(x)\leq V(x),\text{ for all }x\in%
\mathbb{R}
_{+}^{\ast}.
\]

\end{thm}

\begin{pf}
For $\delta>0$ and $\varepsilon>0,$ we denote by $\Phi_{\varepsilon,\delta}$
the function $\Phi_{\varepsilon,\delta}:%
\mathbb{R}
_{+}\times%
\mathbb{R}
_{+}\longrightarrow%
\mathbb{R}
\cup\{-\infty\}$ given by%
\begin{equation}
\Phi_{\varepsilon,\delta}(x,x^{\prime})=U(x)-V(x^{\prime})-\frac
{1}{2\varepsilon}(x-x^{\prime})^{2}-\delta\left(  x^{2}+\left(  x^{\prime
}\right)  ^{2}\right)  , \label{fied}%
\end{equation}
for all $x,x^{\prime}\geq0.$ Suppose that for some $x_{0}\in%
\mathbb{R}
_{+}^{\ast}$ and some $\theta>0$ we have
\[
U(x_{0})-V(x_{0})\geq\theta.
\]
Since $\Phi_{\varepsilon,\delta}$ is upper semi-continuous and $U$ and $V$ are
of linear growth, there exists a global maximum point of $\Phi_{\varepsilon
,\delta},$ denoted by $(x_{\varepsilon,\delta},x_{\varepsilon,\delta}^{\prime
})\in$ $%
\mathbb{R}
_{+}\times%
\mathbb{R}
_{+}.$ Obviously, since $\Phi_{\varepsilon,\delta}(0,x^{\prime})\leq0$ for all
$x^{\prime}\in%
\mathbb{R}
_{+}$, it holds that $x_{\varepsilon,\delta}>0.$ Moreover,
\begin{equation}
\gamma_{\varepsilon,\delta}=\Phi_{\varepsilon,\delta}(x_{\varepsilon,\delta
},x_{\varepsilon,\delta}^{\prime})\geq\Phi_{\varepsilon,\delta}(x_{0}%
,x_{0})\geq\theta-2\delta x_{0}^{2}\geq\frac{\theta}{2}, \label{gamma}%
\end{equation}
for any $\delta\leq\delta_{0}=\frac{\theta}{4x_{0}^{2}}.$ Obviously, for
$\delta\leq\delta_{0}$ fixed, $\left(  \gamma_{\varepsilon,\delta}\right)
_{\varepsilon}$ is increasing and
\[
\gamma_{2\varepsilon,\delta}\geq\gamma_{\varepsilon,\delta}+\frac
{1}{4\varepsilon}\left(  x_{\varepsilon,\delta}-x_{\varepsilon,\delta}%
^{\prime}\right)  ^{2}.
\]
Therefore,
\[
\lim_{\varepsilon\searrow0}\frac{1}{\varepsilon}\left(  x_{\varepsilon,\delta
}-x_{\varepsilon,\delta}^{\prime}\right)  ^{2}=0.
\]
If, for all $\varepsilon>0$ (or, at least for some arbitrary sequence
$\varepsilon_{n}$ such that $\varepsilon_{n}\rightarrow0$ when $n\rightarrow
\infty$) $x_{\varepsilon,\delta}^{\prime}=0,$ then $\lim_{\varepsilon
\searrow0}x_{\varepsilon,\delta}=0,$ and, by taking the upper limit when
$\varepsilon\rightarrow0$ in (\ref{fied}), we get%
\[
\frac{\theta}{2}\leq U(0)-V(0)\leq0,
\]
which contradicts the assumption $\theta>0.$ We deduce that, for
$\varepsilon>0$ small enough, $x_{\varepsilon,\delta}$ and $x_{\varepsilon
,\delta}^{\prime}$ are strictly positive. We consider the test function
\[
\varphi(x)=V(x_{\varepsilon,\delta}^{\prime})+\frac{1}{2\varepsilon}\left(
x-x_{\varepsilon,\delta}^{\prime}\right)  ^{2}+\delta(x^{2}+\left(
x_{\varepsilon,\delta}^{\prime}\right)  ^{2}),\text{ for }x\in%
\mathbb{R}
_{+}^{\ast},
\]
such that $U-\varphi$ has a maximum point at $x_{\varepsilon,\delta}.$ We
write the variational inequality\ and use the previous Lemma to get
\begin{equation}
\max\left\{  H(x_{\varepsilon,\delta},U,\varphi^{\prime}(x_{\varepsilon
,\delta})),1-\varphi^{\prime}(x_{\varepsilon,\delta})\right\}  \geq0.
\label{1}%
\end{equation}
In a similar way we have%
\begin{equation}
\max\left\{  H(x_{\varepsilon,\delta}^{\prime},V,\psi^{\prime}(x_{\varepsilon
,\delta}^{\prime})),1-\psi^{\prime}(x_{\varepsilon,\delta}^{\prime})\right\}
\leq0, \label{2}%
\end{equation}
where
\[
\psi(x^{\prime})=U(x_{\varepsilon,\delta})-\frac{1}{2\varepsilon
}(x_{\varepsilon,\delta}-x^{\prime})^{2}-\delta(x_{\varepsilon,\delta}%
^{2}+(x^{\prime})^{2}),\text{ for all }x^{\prime}\in%
\mathbb{R}
_{+}^{\ast}.
\]

\textbf{(a)} We suppose that
\begin{equation}
H(x_{\varepsilon,\delta},U,\varphi^{\prime}(x_{\varepsilon,\delta}))\geq
H(x_{\varepsilon,\delta}^{\prime},V,\psi^{\prime}(x_{\varepsilon,\delta
}^{\prime})). \label{(a)}%
\end{equation}
Then%
\begin{align}
0  &  \leq\sup_{u\geq\underline{u}}{\Large \{}-r\left(  U(x_{\varepsilon
,\delta})-V(x_{\varepsilon,\delta}^{\prime})\right) \nonumber\\
&  +ap(u)\left[  x_{\varepsilon,\delta}\varphi^{\prime}(x_{\varepsilon,\delta
})-x_{\varepsilon,\delta}^{\prime}\psi^{\prime}(x_{\varepsilon,\delta}%
^{\prime})\right] \nonumber\\
&  +\int_{B}\left(  U(x_{\varepsilon,\delta}-f(ax_{\varepsilon,\delta},y\wedge
u))-V(x_{\varepsilon,\delta}^{\prime}-f(ax_{\varepsilon,\delta}^{\prime
},y\wedge u))\right)  \pi(dy)\nonumber\\
&  +\int_{B}\left(  V(x_{\varepsilon,\delta}^{\prime})-U(x_{\varepsilon
,\delta})\right)  \pi(dy){\LARGE \}.} \label{ineq6}%
\end{align}
We use $\Phi(x_{\varepsilon,\delta};x_{\varepsilon,\delta}^{\prime})\geq
\Phi(x_{\varepsilon,\delta}-f(x_{\varepsilon,\delta},y\wedge u),x_{\varepsilon
,\delta}^{\prime}-f(x_{\varepsilon,\delta}^{\prime},y\wedge u))\ $to get%
\begin{align*}
&  U(x_{\varepsilon,\delta}-f(ax_{\varepsilon,\delta},y\wedge
u))-V(x_{\varepsilon,\delta}^{\prime}-f(ax_{\varepsilon,\delta}^{\prime
},y\wedge u))\\
&  \leq U(x_{\varepsilon,\delta})-V(x_{\varepsilon,\delta}^{\prime})+\frac
{1}{2\varepsilon}\left(  x_{\varepsilon,\delta}-f(ax_{\varepsilon,\delta
},y\wedge u)-x_{\varepsilon,\delta}^{\prime}+f(ax_{\varepsilon,\delta}%
^{\prime},y\wedge u)\right)  ^{2}\\
&  -\frac{1}{2\varepsilon}\left(  x_{\varepsilon,\delta}-x_{\varepsilon
,\delta}^{\prime}\right)  ^{2}\\
&  +\delta\left(  \left(  x_{\varepsilon,\delta}-f(ax_{\varepsilon,\delta
},y\wedge u)\right)  ^{2}-x_{\varepsilon,\delta}^{2}+\left(  x_{\varepsilon
,\delta}^{\prime}-f(ax_{\varepsilon,\delta}^{\prime},y\wedge u)\right)
^{2}-\left(  x_{\varepsilon,\delta}^{\prime}\right)  ^{2}\right) \\
&  \leq U(x_{\varepsilon,\delta})-V(x_{\varepsilon,\delta}^{\prime}),
\end{align*}
and, returning to (\ref{ineq6}), we have%
\begin{align*}
0  &  \leq\sup_{u\geq\underline{u}}\left\{  -r\left(  U(x_{\varepsilon,\delta
})-V(x_{\varepsilon,\delta}^{\prime})\right)  +2ap(u)\left(  \frac{\left(
x_{\varepsilon,\delta}-x_{\varepsilon,\delta}^{\prime}\right)  ^{2}%
}{2\varepsilon}+\delta\left(  x_{\varepsilon,\delta}^{2}+\left(
x_{\varepsilon,\delta}^{\prime}\right)  ^{2}\right)  \right)  \right\} \\
&  \leq\sup_{u\geq\underline{u}}\left(  2ap(u)-r\right)  \times\left(
U(x_{\varepsilon,\delta})-V(x_{\varepsilon,\delta}^{\prime})\right)  .
\end{align*}
Recall that $\sup_{u\geq\underline{u}}2ap(u)\leq\frac{2(1+k_{1})\beta}%
{\zeta_{0}}<r$ (see (A2)). Thus, it follows that $\gamma_{\varepsilon,\delta
}<0$ which contradicts (\ref{gamma}).

\textbf{(b) }If (\ref{(a)}) does not hold, we use (\ref{1}) and (\ref{2}) and
we must have%
\begin{equation}
1-\varphi^{\prime}(x_{\varepsilon,\delta})\geq0\geq1-\psi^{\prime
}(x_{\varepsilon,\delta}^{\prime}), \label{(b)}%
\end{equation}
thus%
\[
\frac{1}{\varepsilon}\left(  x_{\varepsilon,\delta}-x_{\varepsilon,\delta
}^{\prime}\right)  -2\delta x_{\varepsilon,\delta}^{\prime}\geq\frac
{1}{\varepsilon}\left(  x_{\varepsilon,\delta}-x_{\varepsilon,\delta}^{\prime
}\right)  +2\delta x_{\varepsilon,\delta}.
\]
We deduce that $x_{\varepsilon,\delta}^{\prime}=x_{\varepsilon,\delta}=0 $ and
get a contradiction$.$ The proof of the comparison result is now complete.
\end{pf}

\section{Numerical results}

We now turn our attention to some particular case and observe the optimal
retention process by means of numerical simulation. We have seen that, for the
collective risk model introduced in \cite{MS}, a single insurance contract is
considered and, of course, the risk is given for this one contract. A possible
way to extend this model is to suppose that the risk concerns all contracts
(or at least a percentage). We assume that the claims have constant intensity
$\delta$ and the random measure $\mu$ is associated with some Poisson process
of constant intensity $\pi(dy)=\beta G_{\delta}(dy),$ where $G_{\delta}$
corresponds to the Dirac mass. Moreover, the function $f$ is given by
$f(x,y)=\rho xy,$ with $0<\rho\leq1$ (that is only some $\rho$ part of the
total contracts is subject to claims). In this case, the minimal retention
level needed to cover expenditures is given explicitly by $\underline{u}%
=\frac{\left(  k_{2}-k_{1}\right)  \delta}{k_{2}}$ and $p(u)=\left(
k_{1}-k_{2}\right)  \beta\rho\delta+\left(  1+k_{2}\right)  \beta\rho u,$ for
all $\frac{\left(  k_{2}-k_{1}\right)  \delta}{k_{2}}\leq u\leq\delta.$

Under the above assumptions, Eq. (\ref{SDE}) reads%
\begin{equation}
X_{t}^{x,u,L}=x+a\int_{0}^{t}X_{s}^{x,u,L}p(u_{s})ds-\rho a\int_{0}^{t}%
X_{s-}^{x,u,L}u_{s}dN_{s}-\int_{0}^{t}dL_{s}. \label{SDEnum}%
\end{equation}
Theorem 10 states that the maximized expected discounted dividends is the
unique viscosity solution for the Hamilton-Jacobi-Bellman variational
inequality%
\begin{equation}
\left\{
\begin{array}
[c]{l}%
\max\left\{  H(x,V,V^{\prime}(x)),1-V^{\prime}(x)\right\}  =0\text{ in }%
\mathbb{R}
_{+}^{\ast},\\
V(0)=0,
\end{array}
\right.  \label{HJBnum}%
\end{equation}
where%
\[
H(x,V,q)=\sup_{\frac{\left(  k_{2}-k_{1}\right)  \delta}{k_{2}}\leq
u\leq\delta}\left\{  -rV(x)+axp(u)q+\beta\left[  V(x-a\rho xu)-V(x)\right]
\right\}  .
\]
The standard procedure in order to apply numerical arguments is to obtain a
bounded space. Thus, we write the previous equation on $\left[  0,1\right)  $
by taking $y=\frac{x}{x+1}$ and $\psi(y)=V(x).$ This leads to the following
HJB equation%
\begin{equation}
\left\{
\begin{array}
[c]{l}%
\max\left\{  G(y,\psi,\psi^{\prime}(y)),1-\left(  1-y\right)  ^{2}\psi
^{\prime}(y)\right\}  =0\text{ in }\left[  0,1\right)  ,\\
\psi(0)=0,
\end{array}
\right.  \label{HJBnum_bdd}%
\end{equation}
where%
\begin{align*}
&  G(y,\psi,q)=\\
&  \sup_{\frac{\left(  k_{2}-k_{1}\right)  \delta}{k_{2}}\leq u\leq\delta
}\left\{  -r\psi(y)+ap(u)y\left(  1-y\right)  q+\beta\left[  \psi
(\frac{y(1-a\rho u)-a\rho u}{a\rho uy+1-a\rho u})-\psi(y)\right]  \right\}  .
\end{align*}

As in Mnif, Sulem (2005), the approximate solution of \ Eq. (\ref{HJBnum_bdd})
is computed with the help of finite difference approximations and the policy
iteration algorithm.

We consider two particular cases: the first one illustrates the natural
framework in which the reinsurance company perceives a relative safety loading
greater than that of the insurer, while the second example assumes the
opposite. The data set we use is given in the following table%
\[%
\begin{tabular}
[c]{|l|l|l|l|l|l|l|}\hline
& $k_{1}$ & $k_{2}$ & $\delta$ & $r$ & $\beta$ & $\rho$\\\hline
Fig 1 & $0.2$ & $0.25$ & $1$ & $0.07$ & $0.0011$ & $10\%$\\\hline
Fig 2 & $0.2$ & $0.19$ & $1$ & $0.07$ & $0.0011$ & $10\%$\\\hline
\end{tabular}
\]

For the first framework, the optimal retention level turns out to be maximal
as shown by Fig 1.%
\[%
{\parbox[b]{4.862in}{\begin{center}
\includegraphics[
height=3.6538in,
width=4.862in
]%
{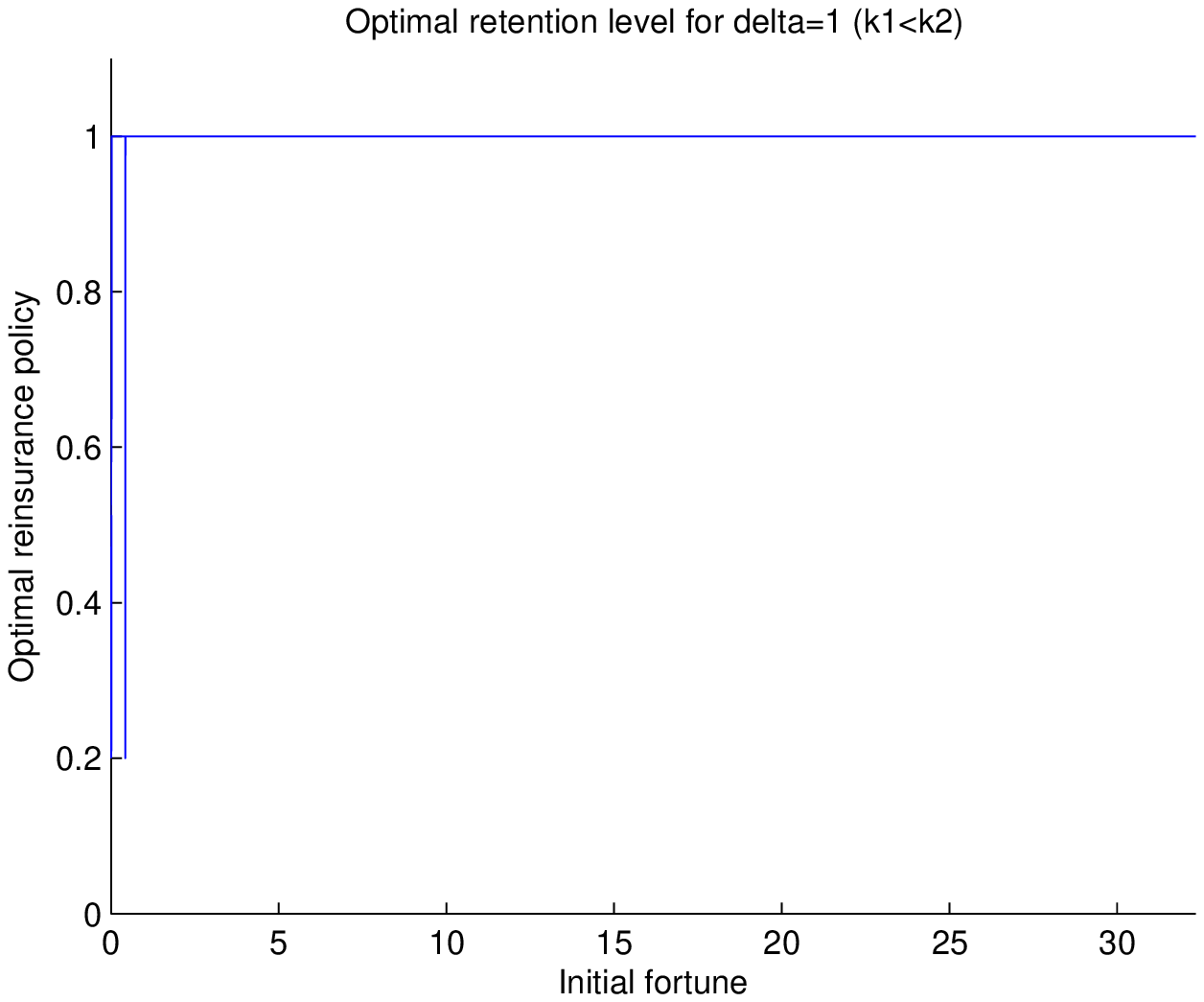}%
\\
Fig 1. Optimal retention level for $\delta=1,$ $k_1=0.2,$ $k_2=0.25$
\end{center}}}
\]

As can be expected in the second case, if the initial reserve is great enough,
then the direct insurer should play the safety card in order to maximize
expected discounted dividends. Indeed, since the relative safety loadings
guarantee a proportional steady income to the insurer, the optimal retention
level is null (see Fig 2).%

\[%
{\parbox[b]{4.8473in}{\begin{center}
\includegraphics[
height=3.6512in,
width=4.8473in
]%
{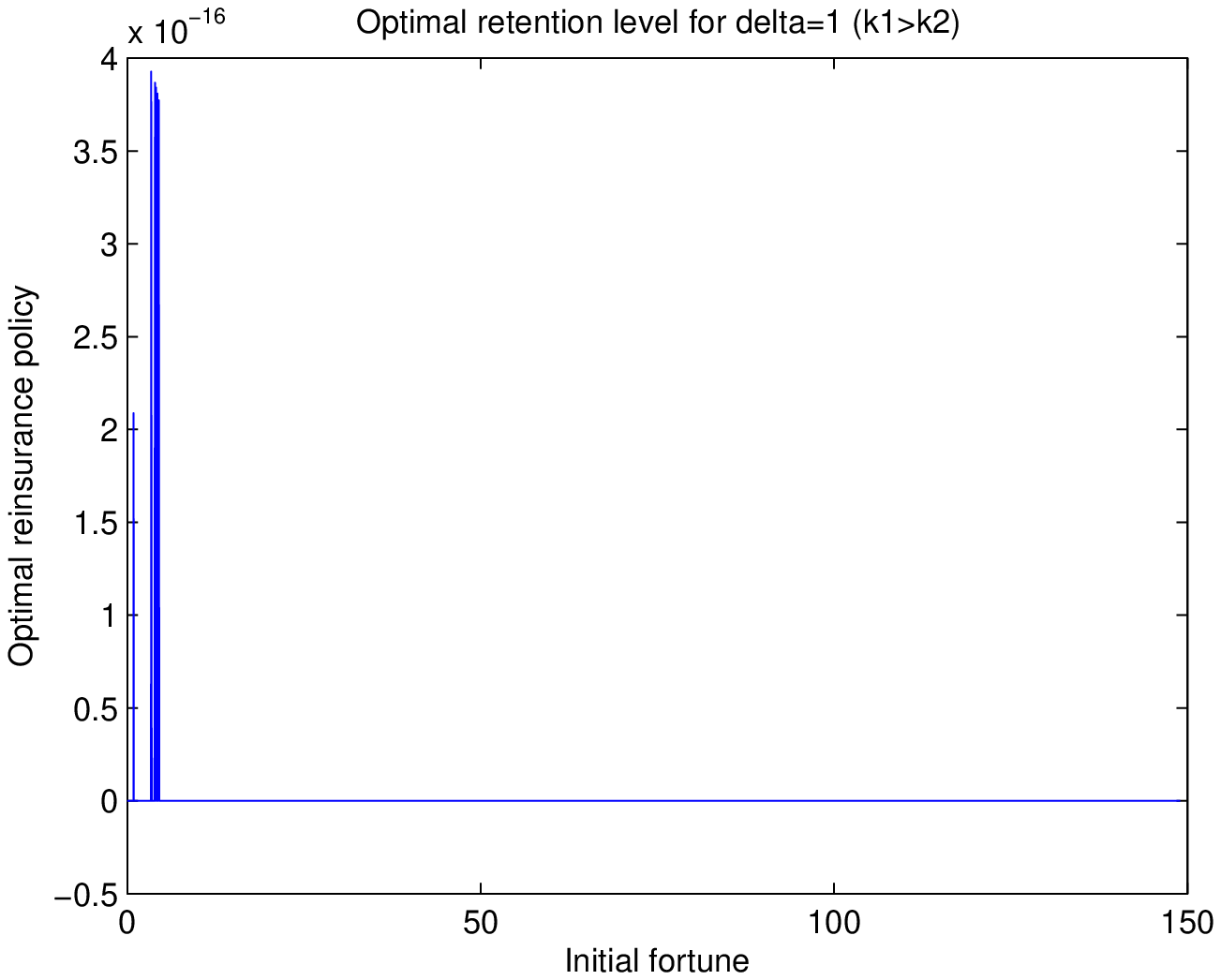}%
\\
Fig 2. Optimal retention level for $\delta=1,$ $k_1=0.2,$ $k_2=0.19$
\end{center}}}
\]

\end{document}